\documentclass[11pt]{article}
\usepackage{setspace}
\usepackage{amssymb}
\usepackage{amsmath}
\usepackage{amsfonts}
\usepackage{slashed}

\def\be{\begin{equation}}

\def\ee{\end{equation}}

\def\bea{\begin{eqnarray}}

\def\eea{\end{eqnarray}}

\newcommand\eps{\epsilon}

\makeatletter
\def\blfootnote{\xdef\@thefnmark{}\@footnotetext}
\makeatother

\begin{document}

\singlespace

\begin{flushright} BRX TH-6709 \\
CALT-TH 2022-028
\end{flushright}

\vspace*{.3in}

\begin{center}

{\Large\bf All identically conserved gravitational tensors are metric variations of invariant actions}

{\large S.\ Deser}

{\it 
Walter Burke Institute for Theoretical Physics, \\
California Institute of Technology, Pasadena, CA 91125; \\
Physics Department,  Brandeis University, Waltham, MA 02454 \\
{\tt deser@brandeis.edu}
}
\end{center}

\begin{abstract}
I prove an old unsolved conjecture, the hard --- necessary --- part of its obvious sufficiency, namely that all identically conserved symmetric $2$-tensors are necessarily metric variations of invariant actions, thus sparing the world more alternative gravity theories. The proof is reasonably simple, if perhaps a ``physicist's". 
\end{abstract}

\section{Introduction}
The possible existence of covariantly conserved gravitational 2-tensors, i.e., those depending only on the metric and curvatures (and their covariant derivatives) that do not stem from an action would seem to have been been raised and solved over a century ago, but to my knowledge was only discussed recently [1]. Given its physical importance --- adding a whole new set of alternate gravity models --- this is a bit surprising, to say the least. Here I fill this gap, showing that they are indeed excluded, at least at a ``physicist's proof" level.

\section{Proof}
We divide all conserved tensors into two disjoint categories, those that do not contain terms with one or more leading covariant derivatives, and those that do. The former consists of two types.  They are, of course, the Einstein tensor $G^{\mu\nu}$ and (in $D>4$) the Lovelock (double density) tensor [2]
\begin{equation*}
L^{\mu\nu} = \eps^{\mu ...} \eps^{\nu ...} g_{..} \,  \, \, g_{..} R_{....}  \, \, R_{....} 
\end{equation*}
Its covariant conservation is manifest, by the Bianchi identities, but what is most relevant is that it is the metric variation of $\int1/\sqrt{-g} g_{\mu\nu} L^{\mu\nu}$ because the metric variation of the curvatures vanishes by Bianchi as well. For Einstein, I need not belabor the same result --- here the variation of the curvature's contribution vanishes since it consists of two covariant derivatives $\sim DD \delta g$, to which the explicit metric in $\int \sqrt{-g} g^{\mu\nu} R_{\mu\nu}$ is transparent. It was proved in [2] that those are the only two conserved tensors containing second derivatives (but no higher) of the metric. This means they have NO explicit $D$ since
they can only act on curvatures, thus making them of higher than 2 derivative order, and subject to part 2. That takes care of part 1.

There is one intermediate, Part 1.5, example, in $D=3$ only, namely, the third derivative Chern-Simons action, whose leading $T$-term is the Cotton tensor's (see [3] for details). Likewise, there is an infinite variety from the basic forms, for example just insert powers of d'Alembertians in the middle of any actions etc. The universal point is that all $T$ must start as projectors by the linearized argument and as can also be seen from the action' s variations --- varying curvatures, even if there are many derivatives on them, the outermost factor is always the projector. The apparent exceptions, Einstein and Lovelock, are not --- it's just that there the projectors annihilate their subjects: see e.g., the first term in (4).

\section{Part 2 --- higher derivatives}
In the linearized limit, where covariant derivatives become ordinary, the most general identically conserved symmetric tensor is $\partial^2_{\alpha\beta} H^{\mu \alpha \nu \beta}$ where $H$ has the algebraic symmetries of the Riemann tensor. 
There are two sub-categories, namely the operator acting on a tensor $h_{\alpha\beta}$ to give linearized Einstein and the still more special $P=(\partial^2_{\mu\nu} - \eta_{\mu\nu} \partial^2)$ acting on a scalar $X$. These are the transverse projectors at linearized level; there, one could also allow all sorts of ordinary derivatives standing before these, but that is not allowed at full level because, as can easily be seen, no $Z$ in (1) can compensate for the ensuing leading $R$ terms in their divergences.

Now back to the full theory. The most general candidate $T^{\mu\nu}$ is the sum of the above three projectors, each acting on its (tensor or scalar) $X$ plus a correction term $Z^{\mu\nu}$. Let me focus on the simplest case --- bad enough --- to show how it goes. We have 
\begin{equation}
T^{\mu\nu}  = P^{\mu\nu} X +Z^{\mu\nu}   
\end{equation}
with the ``scalar" $P$ now made from covariant derivatives; hence the divergence of (1) begins with $R^{\mu\nu} D_{\nu}X$.
What $Z$ can cancel this? It must begin with $R^{\nu\nu} Z'$, whose div is $1/2 R  D_\mu Z' + R^{\mu\nu} D_\nu Z'$. Can this be done? Yes, but we shall see that only if $T$ is action-derived. Let's divide $Z$ into $g^{\mu\nu} Y + R^{\mu\nu} W$
(we could have included a term proportional to Riemann, but it is clearly useless here).
The div of (1) then becomes
\begin{equation}
D_\nu T^{\mu\nu}=0= R^{\mu\nu} D_\nu  X +  R^{\mu\nu} D_\nu  W + 1/2 (D^\mu R) W + D^\mu Y.     
\end{equation}
So $W$ can cancel $X$, but the last two terms must cancel internally. So the $W$ part must be a total divergence, which is possible only if it depends on $R$ alone, but not Ricci or Riemann, as must then $Y$, and hence $X$ as well! That does it: The action and its variations are
\begin{equation}
I= \int \sqrt{-g} F(R),  \, \, \, \, \,  \delta I= \int V\sqrt{-g} \delta g_{\mu\nu}  [(P^{\mu\nu} + R^{\mu\nu}) F' -1/2 g^{\mu\nu} F],             
\end{equation}
in agreement with (2). Note first that the linearized limit is OK, since the explicit $R$ term is higher order, while the last term is absent since the flat metric replaces $\sqrt{-g}$. Also, arbitrary  covariant derivatives inside $F$ are allowed, by following the reasoning below (2). There remains one final term, one that vanishes at linear order, namely $R_{,\mu} R_{,\nu} V(R)$, whose divergence is
\begin{equation}     
(D^2R R_{,\mu} + D_{\mu\nu}R R^{\, \, \,  \nu}_{,}) V + (DR DR) V_{,\mu}.    
\end{equation}
Precisely the same chain that led from (2) to (3) shows that  this term must come from the action 
\begin{equation}
I=  \int \sqrt{-g} R_{...}R_{,\mu} g^{\mu\nu}  R_{,\nu . . . } 
\end{equation}
so this time corrections to $PX$ in (1) are also required.

The moral is that conservation plus linearized limit conservation suffice to constrain any identically conserved $T$ to be the variation of an action, at least for the simplest category (1); strictly speaking, I have not proven this for the slightly more general case where the $R_, R_,$ involve more derivatives and the $V$ is a tensor. This is a physicist's proof: I will not push my luck nor test the reader's patience by proving this for the mixed cases: the reasoning is rather clear and life is short.

This completes our proof: the second category, the one with explicit leading $D$-terms, so of higher derivative order, necessarily begins with what would be transverse projectors at linearized level plus corrections as in our simplest example, and also must stem from an action or else there are no possible $Z$-correction terms.
 
\section{Conclusion}
I have given a physicist's proof that all identically conserved gravity two-tensors are necessary derivable from actions, as has always been tacitly assumed to date. Physicist's because I have only treated one, if generic, case, rather than all three possible, if similar, ones.

\section*{Acknowledgements}
I thank my old collaborator, Y. Pang for useful correspondence, a referee for insisting on better details, and J. Franklin for tech help.  This work was supported by the U.S. Department of Energy, Office of Science, Office of High Energy Physics, Award Number de-sc0011632.

\end{document}